\documentclass[structabstract]{aa}
\usepackage{graphicx}
\usepackage{txfonts}
\usepackage{longtable}

\begin{document}
   \title{A planetary companion around the K giant $\epsilon$ Corona Borealis}
   \author{Byeong-Cheol Lee\inst{1},
          Inwoo Han\inst{1},
          Myeong-Gu Park\inst{2},
          David E. Mkrtichian\inst{3,4},
          \and
          Kang-Min Kim \inst{1}
          }

   \institute{Korea Astronomy and Space Science Institute, 776,
		Daedeokdae-Ro, Youseong-Gu, Daejeon 305-348, Korea\\
	      \email{[bclee;iwhan;kmkim]@kasi.re.kr}
	    \and
	      Department of Astronomy and Atmospheric Sciences,
	      Kyungpook National University, Daegu 702-701, Korea\\
	      \email{mgp@knu.ac.kr}
        \and
          National Astronomical Research Institute of Thailand, Chiang Mai 50200, Thailand
        \and
          Crimean Astrophysical Observatory, Nauchny, Crimea, 98409, Ukraine\\
           \email{davidmkrt@gmail.com}
             }

   \date{Received 5 April 2012 / Accepted 31 August 2012}


  \abstract
   {}
   {Our aim is to search for and study the origin of the low-amplitude and long-periodic radial velocity (RV) variations in K giants.
   }
   {We present high-resolution RV measurements of K2 giant $\epsilon$ CrB from February 2005 to January 2012 using the fiber-fed Bohyunsan Observatory Echelle Spectrograph (BOES) at the Bohyunsan Optical Astronomy Observatory (BOAO).
   }
   {We find that the RV measurements for $\epsilon$ CrB exhibit a periodic variation of 417.9 $\pm$ 0.5  days with a semi-amplitude of 129.4 $\pm$ 2.0 m s$^{-1}$. There is no correlation between RV measurements and chromospheric activity in the Ca II H region, the Hipparcos photometry, or bisector velocity span.
   }
   {Keplerian motion is the most likely explanation, with the RV variations arising from an orbital motion. Assuming a possible stellar mass of 1.7 $\pm$ 0.1 $\it M_{\odot}$ for $\epsilon$ CrB, we obtain a minimum mass for the planetary companion of 6.7 $\pm$ 0.3 $M_{\mathrm{Jup}}$ with an orbital semi-major axis of 1.3 AU and eccentricity of 0.11. We also discuss the implications of our observations for stellar metallicity versus planet occurrence rate and stellar mass versus planetary mass relations.
   }


   \authorrunning{Byeong-Cheol Lee et al.}
   \titlerunning{A planetary companion around the K giant $\epsilon$ Corona Borealis}
   \maketitle
%

\section{Introduction}

As a star evolves towards the red giant branch, its rotational velocity decreases. Simultaneously, the stellar surface begins to cool significantly, and it may develop surface inhomogeneities such as stellar spots. The activity can cause changes in spectral line profiles, which develop in a complex manner with the gravitational influence of a planetary companion. Thus, this and stellar pulsations make it more difficult to detect exoplanets around giants than around main-sequence (MS) stars. Nevertheless, cool evolved stars are suitable samples for precise RV measurements and for studying exoplanets because they offer new clues to help develop theories of planet formation and help explain the long-term survival of planetary systems.

Since the discovery of the first planetary companion around K giant star $\iota$ Dra (Frink et al. 2002), about 30 planetary companions have been detected so far around evolved K giant stars using the precise RV method (Setiawan et al. 2003a, 2003b; Mitchell et al. 2003; Hatzes et al. 2005, 2006; Reffert et al. 2006; Do$\ddot{}$llinger et al. 2007, 2009; Johnson et al. 2008, 2010, 2011; Sato et al. 2008a, 2008b, 2010; de Medeiros et al. 2009; Niedzielski et al. 2009; Han et al. 2010; Lee et al. 2011; Moutou et al. 2011; Wittenmyer et al. 2011; and Gettel et al. 2012). They seem to show different properties than MS stars. Giants that host planets do not favor high-metallicity objects (Pasquini et al. 2007) and trend to harbor more massive planets (Lovis \& Mayor 2007; D$\ddot{o}$llinger et al. 2009).

For the past seven years, we have conducted precise RV measurements of 55 K giants. Here, we present a long-period and low-amplitude RV variation of K giant $\epsilon$ CrB. In Sect. 2, we describe the properties of $\epsilon$ CrB and analyze the periods of RV measurements in Sect. 3. Possible origins of the RV variations are investigated in Sect. 4. Last, we discuss the implications of this work in Sect. 5.


\section{The properties of $\epsilon$ CrB}

The cool evolved star of spectral type of K2 III (ESA 1997), $\epsilon$ CrB (= HD 143107 = HR 5947 = HIP 78159) has an apparent magnitude of 4.14. The main parameters come from Massarotti et al. (2008), who present rotational velocities for a sample of 761 giants selected from the Hipparcos catalog. They measured new rotational velocities for all the sample using spectroscopic line broadening methods with the CfA Digital Speedometers (Latham 1992). They estimated the rotational velocity of $\epsilon$ CrB to be 2.4 km s$^{-1}$. Based on the rotational velocity of 2.4 km s$^{-1}$ and stellar radius of 21 $R_{\odot}$ (Massarotti et al. 2008), we derived the range for the estimated velocity of the rotational period as

   $P_{\rm rot} = 2 \pi R_{\star}$ / ($v_{\rm rot}$ sin $i$) = 442.7 days.\\

\hskip -15pt
The atmospheric parameters of $\epsilon$ CrB were determined based on the program TGVIT (Takeda et al. 2005). We used 245 equivalent width (EW) measurements of Fe I and Fe II lines, which resulted in $T_{\mathrm{eff}}$ = 4406 $\pm$ 15 K, log $\it g$ = 1.94 $\pm$ 0.08, $v_{\mathrm{micro}}$ = 1.59  $\pm$ 0.07 km s$^{-1}$, and $\mathrm{[Fe/H]}$ = -- 0.094 $\pm$ 0.001.

The stellar mass has been determined somewhat differently by individual authors (because it is very sensitive to log $\it g$). We estimated the stellar mass from the theoretical stellar isochrones using its position in the color--magnitude diagram based on Bertelli et al. (1994) and Girardi et al. (2000). We adopted a version of the Bayesian estimation method by J{\o}rgensen \& Lindegren (2005) and da Silva et al. (2006) using the determined values for $T_{\mathrm{eff}}$, $\mathrm{[Fe/H]}$, $M_{v}$, and parallax. Our estimated parameters yield a stellar mass of 1.7 $\pm$ 0.1 $M_{\odot}$. The basic stellar parameters of $\epsilon$ CrB are summarized in Table~\ref{tab1}.


%
\begin{table}
\begin{center}
\caption[]{Stellar parameters for $\epsilon$ CrB.}
\label{tab1}
\begin{tabular}{lcc}
\hline
\hline
    Parameter          & Value      &    Reference     \\

\hline
    Spectral type            & K2 III    & Hipparcos  \\
    $\textit{$m_{v}$}$ [mag] & 4.14      & Hipparcos  \\
    $\textit{B-V}$ [mag]     & 1.23      & Hipparcos  \\
    age [Gyr]                & 1.74 $\pm$ 0.37\tablefootmark{a} & Derived  \\
    Distance [pc]            & 67.9        & Shaya \& Olling (2011)    \\
    RV [km s$^{-1}$]         & -- 30.92      & Massarotti et al.  (2008) \\
    Parallax [mas]           & 13.97 $\pm$ 0.68  & Kharchenko et al. (2009) \\
    Diameter [mas]           & 3.08 $\pm$ 0.16\tablefootmark{b} & Richichi \& Percheron (2002) \\
    $T_{\mathrm{eff}}$ [K]   & 4365 $\pm$ 9      & Massarotti  et al. (2008) \\
                             & 4406  $\pm$ 15    & This work                 \\
    $\mathrm{[Fe/H]}$        & -- 0.32           & Massarotti  et al. (2008) \\
                             & -- 0.094 $\pm$ 0.001  & This work      \\
    log $\it g$              & 2.3                & Massarotti et al. (2008) \\
                             & 1.94 $\pm$ 0.08    & This work \\
    $\textit{$R_{\star}$}$ [$R_{\odot}$] & 21                & Massarotti  et al. (2008) \\
    $\textit{$M_{\star}$}$ [$M_{\odot}$] & 1.7 $\pm$ 0.1\tablefootmark{a}  & Derived     \\
    $\textit{$L_{\star}$}$ [$L_{\odot}$]      & 151               & Massarotti  et al. (2008) \\
    $v_{\mathrm{rot}}$ sin $i$ [km s$^{-1}$]  & 2.4               & Massarotti  et al. (2008) \\
    $P_{\mathrm{rot}}$ / sin $i$ [days]       & 442.7             &  Derived                  \\
    $v_{\mathrm{micro}}$ [km s$^{-1}$]        & 1.59  $\pm$ 0.07  &  This work   \\

\hline

\end{tabular}
\end{center}
\tablefoottext{a}{Derived using the online tool (http://stevoapd.inaf.it/cgi-bin/param}).
\tablefoottext{b}{Computed on the basis of a formula as described in method, and error is included whenever possible. The uniform disk angular diameter is  2.73 $\pm$ 0.03 mas, and the limb-darkened angular diameter is 2.80 $\pm$ 0.03 mas.}
\end{table}
%


\section{Observations and analysis}

Since 2003, we have conducted a precise RV survey of 55 K giants using the fiber-fed high-resolution ($\emph{R}$ = 90 000) the Bohyunsan Observatory Echelle Spectrograph (BOES; Kim et al. 2007) attached to the 1.8-m telescope at Bohyunsan Optical Astronomy Observatory (BOAO) in Korea.

We acquired 52 spectra for $\epsilon$ CrB from February 2005 to January 2012. Each estimated signal-to-noise ratio (S/N) at an iodine (I$_{2}$) wavelength region (4900 -- 6000 {\AA}) is about 250 with a typical exposure time ranging from 240 to 480 seconds. An I$_{2}$ absorption cell was used to provide precise RV measurements. The extraction of normalized 1--D spectra was carried out with IRAF (Tody 1986) software and DECH (Galazutdinov 1992) code. The I$_2$ analyses and precise RV measurements were undertaken using the RVI2CELL (Han et al. 2007) developed at the Korea Astronomy $\&$ Space Science Institute (KASI). The RV measurements for $\epsilon$ CrB are listed in Table~\ref{tab2}.

The RV standard star $\tau$ Ceti shows a stable RV with an rms scatter of 6.8 m s$^{-1}$ (Figure~\ref{orbit}) over the time span of our observations and demonstrates the long-term stability of the BOES (Lee et al. 2011). RV measurements of $\epsilon$ CrB show a standard deviation of 76.4 m s$^{-1}$, which is 11 times more than for the RV standard star (Figure~\ref{orbit}).

Period analysis is applied in order to determine the significance of periodic trends or variabilities in the time series. The Lomb-Scargle periodogram (Lomb 1976; Scargle 1982), a useful tool for investigating long-period variations for unequally spaced data, was calculated for the RV time series for $\epsilon$ CrB. Figure~\ref{power1} (top panel) shows a significant power at $f_{1}$ = 0.00239 c\,d$^{-1}$ ($P$ = 417.9 days).

A false alarm probability (FAP) is a metric that expresses the significance of a period. A false alarm arises in period analysis techniques when a period is incorrectly found where none exists in reality. We determined the significance of a period by calculating an FAP for the dominant period by a bootstrap randomization technique (Press et al. 1992; K\"{u}rster et al. 1999). We found an FAP of less than $10^{-5}$ for the period of 418 days. Figure~\ref{power1} (bottom panel) shows the Lomb-Scargle periodogram of the residual after subtracting the frequency of $f_{1}$ = 0.00239 c\,d$^{-1}$, and we were able to find a remaining peak at the frequency of $f_{1}$ = 0.00497 c d$^{-1}$ ($P$ $\sim$ 201 days) with an FAP of 1.4 $\pm$ 0.2\%.

%
\begin{table}
\begin{center}
\caption{RV measurements for $\epsilon$ CrB from February 2005 to January 2012 using the BOES.}
\label{tab2}
\begin{tabular}{cccccc}
\hline\hline

 JD         & $\Delta$RV  & $\pm \sigma$ &        JD & $\Delta$RV  & $\pm \sigma$  \\
 -2 450 000 & m\,s$^{-1}$ &  m\,s$^{-1}$ & -2 450 000  & m\,s$^{-1}$ &  m\,s$^{-1}$  \\
\hline

3430.324001  &    -81.2 &   6.7 &   4994.182859 &   -48.5 &  17.8 \\
3460.235768  &      2.5 &   7.4 &   4995.208262 &   -45.3 &   6.9 \\
3460.297588  &     -9.7 &   7.0 &   4996.099283 &   -36.8 &   7.0 \\
3506.253786  &    130.3 &   8.2 &   4996.106632 &   -23.7 &   9.4 \\
3545.042399  &    123.5 &   7.3 &   5130.932411 &    -4.1 &   8.3 \\
3730.408100  &    -65.3 &   8.1 &   5248.368418 &   173.7 &   6.6 \\
3818.237686  &    -68.4 &   6.2 &   5321.137848 &   144.6 &   6.8 \\
3821.138258  &    -80.1 &   7.3 &   5356.148060 &    48.7 &   7.9 \\
3891.145666  &     43.3 &   8.6 &   5810.927302 &    50.3 &   7.1 \\
3899.184958  &     60.4 &   7.3 &   5810.993537 &    27.4 &   6.7 \\
4147.298612  &    -57.9 &   6.7 &   5820.022432 &   -61.3 &   6.8 \\
4210.246254  &    -83.0 &   9.2 &   5820.028150 &   -62.0 &   7.3 \\
4262.234769  &   -100.3 &   8.7 &   5820.034017 &   -64.2 &   8.0 \\
4264.223335  &    -25.9 &   7.3 &   5841.950835 &   -53.2 &   6.6 \\
4458.409299  &    150.6 &   6.9 &   5841.955858 &   -52.4 &   6.7 \\
4483.406189  &    158.4 &   8.7 &   5841.960546 &   -50.3 &   6.6 \\
4505.293538  &    113.9 &   8.0 &   5842.942221 &   -54.5 &   7.0 \\
4516.321792  &     86.8 &   7.8 &   5842.942221 &   -54.5 &   7.0 \\
4536.294962  &     17.9 &   7.5 &   5842.942221 &   -54.5 &   7.0 \\
4538.204751  &     33.1 &   7.8 &   5911.396383 &   -33.0 &  16.8 \\
4619.213839  &   -138.5 &  18.8 &   5911.401059 &   -25.3 &  20.0 \\
4847.360348  &    149.9 &   8.0 &   5911.406117 &   -50.1 &  18.8 \\
4930.186145  &     83.3 &   6.9 &   5933.351526 &   -23.0 &   8.8 \\
4970.217174  &     -5.5 &   8.5 &   5933.355508 &   -17.1 &   8.9 \\
4970.224605  &    -10.7 &   8.1 &   5933.359501 &   -26.2 &   7.9 \\
4971.157049  &     -6.9 &   7.3 &   5933.363483 &   -26.9 &   7.7 \\

\hline

\end{tabular}
\end{center}
\end {table}


\section{Origin of the RV variations}

RV of giant stars may be affected by surface phenomena and by the presence of an orbiting planet. While short-term RV variations (hours to days) with a low-amplitude in evolved stars have been known to be the result of stellar pulsations (Hatzes \& Cochran 1998; Hekker et al. 2006), long-term RV variations (hundreds of days) may be caused by three kinds of phenomena:
stellar pulsations, rotational modulations by inhomogeneous surface features, or planetary companions.

\subsection{Orbital mode}

\begin{figure}
   \centering
   \includegraphics[width=8cm]{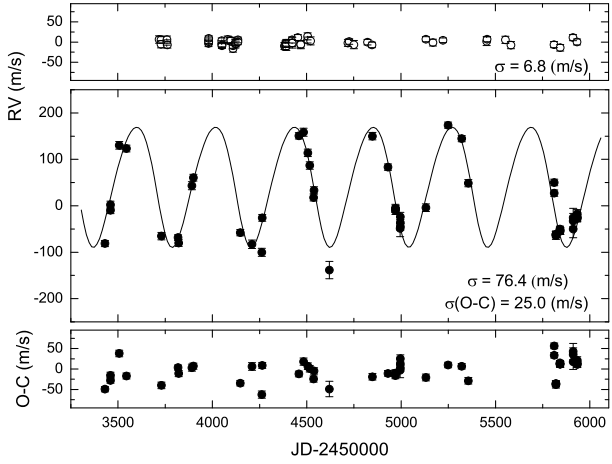}
      \caption{Variations in the RV standard $\tau$ Ceti (\emph{top panel}), RV curve (\emph{middle panel}), and rms scatter of the residual (\emph{bottom panel}) for $\epsilon$ CrB from February 2005 to January 2012. The solid line is the orbital solution with a period of 418 days and an eccentricity of 0.11.
              }
         \label{orbit}
   \end{figure}

\begin{figure}
   \centering
   \includegraphics[width=8cm]{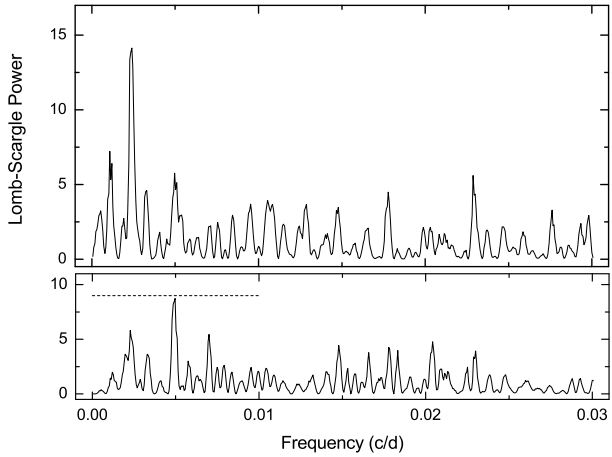}
      \caption{The Lomb-Scargle periodogram of the RV measurements for $\epsilon$ CrB.
      The periodogram shows a significant power at a frequency of 0.00239 c d$^{-1}$
      corresponding to a period of 417.9 days (\emph{top panel}). And after subtracting the main frequency variations (\emph{bottom panel}), the highest significant power at a frequency is 0.00497 c d$^{-1}$ ($P$ = 201.2 days). The horizontal dashed line indicates an FAP threshold of 1\%.
      }
         \label{power1}
   \end{figure}

The source $\epsilon$ CrB reveals a periodic signal in our RV variations, and we find the best-fit Keplerian orbit with a $P$ = 417.9 $\pm$ 0.5 days, a semi-amplitude $K$ = 129.4 $\pm$ 2.0 m s$^{-1}$, and an eccentricity $e$ = 0.11 $\pm$ 0.03. Figure~\ref{orbit} shows the RV curve as a function of time for $\epsilon$ CrB and the residual after extracting the main frequency. As can be seen in Figure~\ref{orbit}, the rms of the RV residuals are 25 m s$^{-1}$, which is significantly larger than the RV precision of 6.8 m s$^{-1}$ and than the typical internal error in individual $\epsilon$ CrB RV accuracy ($\sim$ 8 m s$^{-1}$). There are no obvious peaks in the residual periodogram (Figure~\ref{power1}). It is a general tendency that the rms of RV of a median value of 20 m s$^{-1}$ is shown in K giants (Hekker et al. 2006). We derive the minimum mass of a planetary companion $m$ sin $i$ = 6.7 $\pm$ 0.3 $M_{\mathrm{Jup}}$ at a distance of $a$ = 1.3 AU from $\epsilon$ CrB. All the orbital elements are listed in Table~\ref{tab3}.

%
%

%
\subsection{Chromospheric activity}

\begin{figure}
   \centering
   \includegraphics[width=8cm]{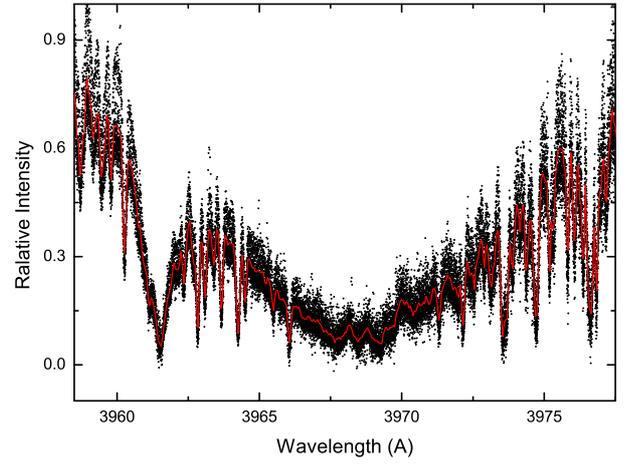}
      \caption{Ca II H spectral region for $\epsilon$ CrB. The dotted lines are individual spectra, and the red solid line is the mean spectrum.
        }
        \label{Ca1}
   \end{figure}

The EW variations of Ca II H \& K lines are sensitive to atmospheric activity and are frequently used as chromospheric activity indicators. The emissions in the Ca II H \& K core are formed in the chromosphere and show a typical central reversal in the chromospheric activity (Pasquini et al. 1988; Saar \& Donahue 1997). The existence of extra emission at the center of the line implies that the source function in the chromosphere is greater than in the photosphere. Unfortunately, because the $\epsilon$ CrB data do not have a sufficient S/N in the Ca II K line region, we used the Ca II H line region, which has a bit higher S/N than does the Ca II K line region. Although the S/N is insufficient, the Ca II H line is enough to estimate the emission feature in the line core, and it does not show any visible emission features. Among the 52 samples of Ca II H line, 35 spectra with relatively high S/N were drown for comparison. Figure~\ref{Ca1} shows individual features in the Ca II H line region, and the red line is the mean spectrum. Although there is no prominent emission in the central region, it exhibits some scatter, which could be from either noise or chromospheric variations.

\begin{figure}
   \centering
   \includegraphics[width=8cm]{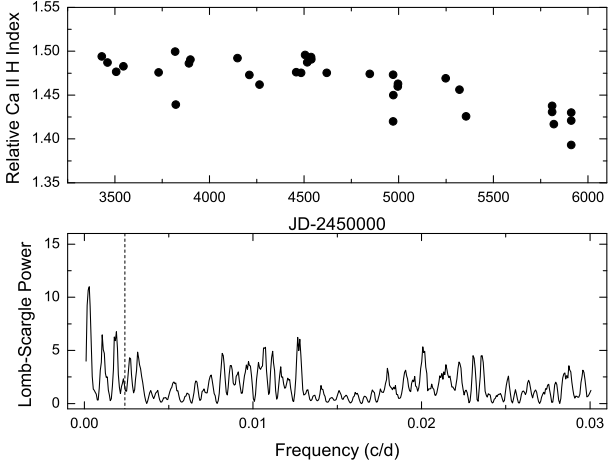}
      \caption{The examinations of the origin of the chromospheric activity (Ca II H line) for $\epsilon$ CrB. It shows JD vs. Ca II H EW variations (\emph{top panel}) and the Lomb-Scargle priodogram of the variations (\emph{bottom panel}). The vertical dashed line marks the location of the periods of 418 days.
        }
        \label{Ca2}
   \end{figure}

We also measured the variations in the Ca II H line EW, which was estimated in a spectral range of 3967.7 -- 3969.3 ${\AA}$ centered on the core of the Ca II H line to avoid nearby blending lines. The mean EW is measured to be 1464.0 $\pm$ 27.8 m${\AA}$. The rms of 27.8 m${\AA}$ corresponds to a 1.9\% variation in the EW. Figure~\ref{Ca2} shows Ca II H EW variations as a function of time and the Lomb-Scargle periodogram. There is significant power at the frequency of 0.00028 c d$^{-1}$, corresponding to a period of 3579 days, which is unrelated to the RV period of 418 days.

\subsection{Hipparcos photometry}

\begin{figure}
   \centering
   \includegraphics[width=8cm]{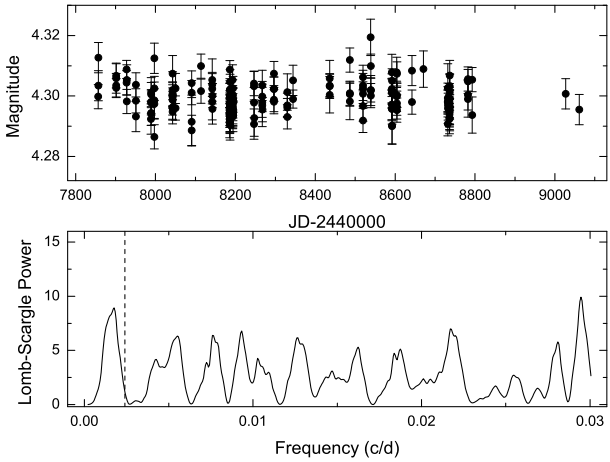}
      \caption{The examinations of the origin of the Hipparcos photometric variations for $\epsilon$ CrB. It shows the Hipparcos photometric measurements (\emph{top panel}) from November 1989 to March 1993 and the Lomb-Scargle priodogram of the variations (\emph{bottom panel}). The vertical dashed line marks the location of the periods of 418 days.
        }
        \label{Hip}
   \end{figure}

We analyzed the photometry to search for possible brightness variations due to the rotational modulation of cool stellar spots or due to the stellar pulsations. The photometry database available for $\epsilon$ CrB comes from 180 Hipparcos measurements from November 1989 to March 1993. The data maintained a photometric stability down to an rms scatter of 0.0066 magnitude corresponding to 0.15\% variations over the time span of the observations. Figure~\ref{Hip} shows the Hipparcos photometric variations and the Lomb-Scargle periodogram of the measurements. There are no significant peaks near the period of 418 days.

%
%
%

%
%
%

%
\subsection{Line bisector variations}

Stellar rotational modulations by surface features can create variable asymmetries in the spectral line profiles,  as well as RV variations. Thus, the variations in the shapes of spectral lines may identify the origin of RV variations. The difference in the width of the line at the top and at the bottom of the line profile is defined as bisector velocity span (BVS).

\begin{figure}
   \centering
   \includegraphics[width=8cm]{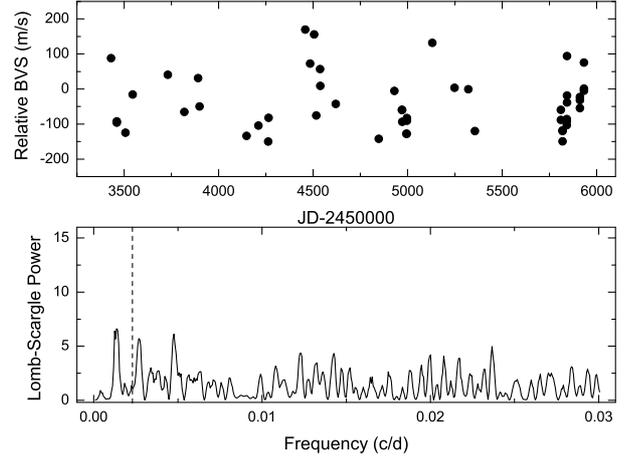}
      \caption{The examinations of the origin of the BVS variations for $\epsilon$ CrB.
      It shows JD vs. BVS variations (\emph{top panel}) from February 2005 to January 2012 and the Lomb-Scargle priodogram of the BVS variations (\emph{bottom panel}).
      The vertical dashed line marks the location of the periods of 418 days.
        }
        \label{BVS}
   \end{figure}

We measured the BVS using the least-squares deconvolution (LSD) technique (Donati et al. 1997; Reiners \& Royer 2004;  Glazunova et al. 2008), which is calculated by the mean profile of the spectral lines. We also used the Vienna Atomic Line Database (VALD; Piskunov et al. 1995) to prepare the list of comparison lines. A total of $\sim$ 3000 lines within the wavelength region of 4500 -- 4900 {\AA} were used to construct the LSD profile. It excluded spectral regions around the I$_{2}$ absorption region (4900 -- 6000 {\AA}), hydrogen lines, and regions with strong contamination by terrestrial atmospheric lines. Finally, we estimated the BVS of the mean profile between two different flux levels of central depth levels of 0.8 and 0.25 as the span points.
BVS variations as a function of time and the Lomb-Scargle periodogram of the BVS for $\epsilon$ CrB are shown in Figure~\ref{BVS}. There are no obvious periodic variations in the BVS measurements.

%
%
%

%
\begin{table}
\begin{center}
\caption{Orbital parameters of the RV measurements for $\epsilon$ CrB.}
\label{tab3}
\begin{tabular}{lc}
\hline
\hline
    Parameter                            & Value                             \\

\hline
    Period [days]                        & 417.9  $\pm$ 0.5                  \\
    $\it T$$_{\mathrm{periastron}}$ [JD] & 2451235.3 $\pm$ 39.7 (1999.152)   \\
    $\it{K}$ [m s$^{-1}$]                & 129.4  $\pm$ 2.0                  \\
    $\it{e}$                             & 0.11   $\pm$ 0.03                 \\
    $\omega$ [deg]                       & 133.1   $\pm$ 33.3                \\
    $f(m)$ [$\it M_{\odot}$]             & (9.227) $\times$ 10$^{-8}$        \\
    $a$ sin $i$ [AU]                     & (4.943) $\times$ 10$^{-3}$        \\
    $\sigma$ (O-C) [m s$^{-1}$]          & 25.0                              \\
\hline
    with $\textit{$M_{\star}$}$ = 1.7 $\pm$ 0.1 [$M_{\odot}$]   &            \\
    $m$ sin $i$ [$M_{\mathrm{Jup}}$]     & 6.7 $\pm$ 0.3                     \\
    $\it{a}$ [AU]                        & 1.3                               \\
\hline

\end{tabular}
\end{center}
\end{table}
%


\section{Discussion}

From the analysis of the seven-year precise RV measurements for $\epsilon$ CrB, we found strongly periodic variability in RV measurements with a period of 418 days and a semi-amplitude of 129 m s$^{-1}$. There is no correlation between the RV measurements and the chromospheric activity, the Hipparcos photometry, or the BVS. Thus, RV variations caused by a planetary companion seem the most likely explanation for the observed RV variations.

We considered two interesting questions related to planets in evolved stars. First, a stellar chemical composition appears to be a major indicator of its probability of hosting planets. MS stars hosting planets are metal-rich (Santos et al. 2004; Fischer \& Valenti 2005), and a survey of 160 metal-poor MS stars give no evidence of exoplanets (Sozzetti et al. 2009). In contrast to MS stars, giants hosting planets follow the same age-metallicity distribution as giants without planets (Pasquini et al. 2007). Recently, unlike the claim of Pasquini et al. (2007), Quirrenbach et al. (2011) suggest that there is a clear trend for the planetary companion frequency to increase with metallicity just as with MS stars. However, since the sample selection criteria might have a significant influence on the correlations between companion frequency and metallicity in addition to the not having enough giants known so far, it would be too premature to reach a conclusion.

\begin{figure}
   \centering
   \includegraphics[width=8cm]{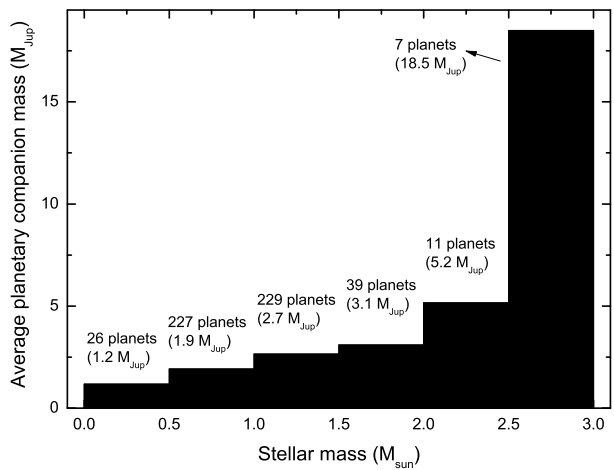}
      \caption{Average mass of planetary companions as a function of stellar mass, taking all planets known as of February 2012 into account.
              }
         \label{stats}
   \end{figure}

Second, although the sample of intermediate-mass stars hosting planets is still limited, there could be a general tendency between the stellar mass and planetary companions being displayed. It gives useful information on the yield of the planetary formation process. We consider six bins of stellar masses with equally spaced stellar mass of 0.5, up to 3 $M_{\odot}$, above which samples are too scarce, and compute an average mass of planetary companions in each bin. Figure~\ref{stats} shows the results in a histogram, where more massive stars do form more massive planetary companions (Johnson et al. 2007; Lovis \& Mayor 2007; D{\"o}llinger et al. 2009), and an average planetary companion mass clearly increases in the range of stars with more than 2.5 stellar mass.
Massive planetary or brown-dwarf mass objects originate in the circumstellar disk, as for giant exoplanets (Niedzielski et al. 2009). In principle, such a system could form from a sufficiently massive protoplanetary disk by means of the standard core-accretion mechanism (Ida \& Lin 2005). Because more massive stars probably have more massive disks and higher surface densities, larger amounts of matter can be accreted; thus, the frequency of massive planets could be higher. With 1.7 $M_{\odot}$ and a companion of 6.7 $M_{\mathrm{Jup}}$, $\epsilon$ CrB seems to be consistent with this speculation. However, there is the obvious observational bias that it is more difficult to detect low-mass planetary companions around a more massive star. Therefore, confirmation of this suggestion requires more discoveries of systems like $\epsilon$ CrB.

There has been an attempt to link larger disk masses of more massive stars to a possibility that the frequency of planets around giants is less dependent on metallicity than is the case for dwarf stars (Ghezzi et al. 2010). In that work, the lower metallicity for the giants hosting planets in comparison to the dwarfs hosting planets is construed as related to the stellar mass. In the core-accretion mechanism of planet formation, higher disk masses can contain an amount of metals available to form giant planets even at lower metallicities.

%


\begin{acknowledgements}
      BCL acknowledges partial support by the KASI (Korea Astronomy and Space Science Institute) grant 2012-1-410-03. Support for MGP was provided by the National Research Foundation of Korea to the Center for Galaxy Evolution Research. DEM acknowledges his work as part of the research activity of the National Astronomical Research Institute of Thailand (NARIT), which is supported by the Ministry of Science and Technology of Thailand. We thank the developers of the Bohyunsan Observatory Echelle Spectrograph (BOES) and all staff of the Bohyunsan Optical Astronomy Observatory (BOAO). This research made use of the SIMBAD database, operated at the CDS, Strasbourg, France.

\end{acknowledgements}
%


\end{document}